\documentclass[pra, twocolumn, superscriptaddress]{revtex4-1}
\usepackage{amsmath}
\usepackage{amssymb}
\usepackage{graphicx}
\usepackage{dcolumn}
\usepackage{braket}
\usepackage{bm}
\usepackage{ulem}
\usepackage{setspace}
\usepackage[colorlinks=true, citecolor=blue, urlcolor=blue, linkcolor=blue ]{hyperref}
\usepackage{multimedia}
\usepackage{bbm}
\hypersetup{breaklinks=true}

\begin{document}\large   

\title{Entanglement resonance in the asymmetric quantum Rabi model}
\author{Yu-Qing Shi}
\affiliation{School of Physical Science and Technology, Lanzhou University, 
Lanzhou 730000, China}
\affiliation{College of Electrical Engineering, Northwest Minzu University, 
Lanzhou 730030, China}
\author{Lei Cong}\email{congllzu@gmail.com}
\affiliation{International Center of Quantum Artificial Intelligence for 
Science and Technology (QuArtist) \\ 
and Department of Physics, Shanghai University, 200444 Shanghai, China}
\affiliation{Helmholtz-Institut, GSI Helmholtzzentrum fur Schwerionenforschung, Mainz 55128, Germany}
\affiliation{Department of Applied Physics, Nanjing Tech University, Nanjing 210009, China}
\author{Hans-Peter Eckle}\email{hans-peter.eckle@uni-ulm.de}
\affiliation{Humboldt Study Centre, Ulm University, Ulm D-89069, Germany}

\begin{abstract}
We investigate the entanglement features in the interacting system of a quantized optical 
field and a two-level system which is statically driven, 
known as the asymmetric quantum Rabi model (AsymQRM). 
Intriguing entanglement resonance valleys with the increase of the 
photon-atom coupling strength and peaks with the increase of the driving 
amplitude are found. 
It is revealed that both of these two kinds of entanglement resonance are caused 
by the avoided level crossing of the associated eigenenergies. 
In sharp contrast to the quantum Rabi model, the entanglement of the AsymQRM 
collapses to zero in the strong coupling regime except when the driving amplitude 
equals to $m\omega/2$, with $m$ being an integer and $\omega$ being the photon 
frequency. 
Our analysis demonstrates that such entanglement reappearance is induced by the hidden 
symmetry of the AsymQRM. 
Supplying an insightful understanding on the AsymQRM, our 
results will be helpful in exploring the hidden symmetry and in 
preparing photon-atom 
entanglement in light-matter coupled systems.
\end{abstract}


\maketitle
\section{Introduction}

Light-matter interaction is described by coupling Bosonic and Fermionic subsystems.
Modelling the interaction between Bosonic
modes
and the electrons of atomic levels
plays therefore a fundamental role in the physics
of strongly interacting quantum systems \cite{HarocheRaimond2006},
especially, but not exclusively,
 in models related to Quantum Optics
 \cite{GarrisonChiao2008}.\\
\indent
For instance, {\it resonance phenomena} of
optical systems where the frequency of a single-mode (Bosonic) radiation 
field
is such that it
couples approximately only to two relevant atomic levels, i.e. a
two-level system (TLS) or qubit (see below for this latter designation), have attracted
considerable interest  since many years 
\cite{AllenEberly1987}
and continue to do so, as evidenced in many recent publications,
a pertinent selection of which we are going to reference
in the course of this publication, especially its introduction.
The two-level system can be described also by spin degrees of freedom.
\\
\indent
While the semi-classical treatment of such systems 
\cite{Rabi}
by I. I. Rabi 
initiated the study of the eponymous model, 
the quantum version, introduced by Jaynes and Cummings \cite{JC}
(for more details on the quantum Rabi model (QRM), see e.g. 
\cite{Book8.10}),
has attracted considerable recent interest. 
A theoretical breakthrough was achieved by the exact solution of the QRM
\cite{DB2011,ChenQingHu2012} 
using two different methods,
which, however, both crucially used the $\mathbb{Z}_2$ symmetry of the QRM
and a Bargmann space analysis
(for these and further recent theoretical developments, see 
\cite{YuEtAl,DB2013} 
and,
\cite{DB2015},
and for recent reviews
\cite{sym11101259,XieQiongtao2017,AlexandreReview,LarsonBook,ArrazolaThesis}
).\\
\indent
The QRM has been realized experimentally in solid state devices,
physical systems
including 
cavity quantum electrodynamic (cavity-QED) 
\cite{HarocheRaimond2006},
circuit quantum electrodynamic (circuit-QED) 
\cite{circuitQEDReview2021}
systems,
and also in trapped ion systems
\cite{PhysRevX.8.021027,CaiEtAl2021}.\\
\indent
In recent years, it became possible, triggered by the technological possibilities these 
and further experimental systems have opened up, to tune
the parameters of the QRM, in particular to reach new regimes where the interaction
between the Bosonic modes and the TLS is strong, i.e. 
to reach the so-called ultra-strong (where the ratio of the field-TLS coupling strength
$g$ and the field frequency $\omega$ is between $0.1$ and $1$)
\cite{Gnter2009,Niemczyk2010,PhysRevLett.105.237001,Yoshihara}
and even the so-called deep strong ($g/\omega>1$) coupling regimes,
predicted theoretically
\cite{PhysRevLett.105.263603}
and realized experimentally in photonic systems 
\cite{PhysRevLett.108.163601}.
Through these developments the necessity arose to consider fully the QRM 
(for a recent review, see
\cite{SCReview2019})
instead of the
simpler quantum Jaynes-Cummings model which is obtained from the QRM by applying
the rotating wave approximation (RWA)
\cite{JC}.
\\
\indent
Further intriguing developments 
include the theoretical prediction of a few-body quantum phase transition in the QRM,
\cite{HPP,PHCP}
which has subsequently been observed experimentally in a single trapped ion
\cite{CaiEtAl2021},
exited state quantum phase transitions 
\cite{PHP} and
quantum phase transitions in extensions of the QRM that include symmetry breaking 
terms,
the focal point of the present study, and
non-linear interaction terms
\cite{PhysRevA.103.063701}
and the application of the QRM in quantum metrology
\cite{QMetro}.
\\
\indent
{
Recently systems related to the QRM have been studied especially in connection with 
quantum computing
\cite{NakaharaOhmi2008,SuterStolze2004,Nielsen2010}
where the fundamental building units (qubits) are two-level systems. 
Moreover, many physically interesting generalizations of the QRM have been
examined (for a recent review of the theoretical developments in this
area, see 
\cite{XieQiongtao2017}).\\
\indent
In this paper,
we shall especially consider the asymmetric extension of the QRM (AsymQRM)
\cite{DB2011},
where a 
$\mathbb{Z}_2$ symmetry breaking term, a driving term, 
$\epsilon\sigma_x$, is added to the Hamiltonian of the QRM.
Other than in cavity QED systems, such 
a term arises naturally in the solid
state devices mentioned earlier
\cite{Gnter2009,Niemczyk2010,PhysRevLett.105.237001,Yoshihara}.
The AsymQRM has earlier been
proposed for experimental realization in 
a micromechanical resonator coupled to a Cooper-pair box
\cite{Armour2002,PhysRevB.68.155311}.\\
\indent
The AsymQRM 
is of great theoretical importance,
especially because the broken $\mathbb{Z}_2$ symmetry is restored for particular values
of the diving amplitude $\epsilon$ 
\cite{DB2011}
and is currently positioned at the centre of the fundamental quest
for a characterisation of integrability in the quantum regime
\cite{Larson2013}.\\
\indent
Before we return to the focus of this paper, the asymmetric QRM, we 
mention in passing
one of the numerous other extensions of the QRM that
are currently under intense investigation and
which originated in a suggestion for an experimental arrangement creating
a non-linear interaction between the Bosonic modes and the TLS
\cite{GP}.
This so-called Stark-term, $\gamma\sigma_z a^\dagger a$,
preserves the $\mathbb{Z}_2$ symmetry of the QRM and
has also been rigorously solved using a Bargmann representation 
\cite{Poles,EJ2017,XieDuanChenStark}
and its physical properties been further investigated
\cite{cong2020selective}.
This model is also discussed as a promising candidate for another foray to shed light on
the fundamental notion of quantum integrability \cite{EckleUnpub}.
The reason for this expectation is that,
within the rotating wave approximation, 
the quantum Jaynes-Cummings model 
admits an exact solution 
using the infinite Lie algebra approach of
the Richardson-Gaudin type Bethe ansatz
\cite{RGBAReview,GaudinCaux,Book12}
while adding a non-linear Stark term $\gamma a^\dagger a\sigma_z$ 
renders the quantum Jaynes-Cummings-Stark
model 
amenable to an
algebraic Bethe ansatz and the model is thus a Yang-Baxter integrable model
\cite{Book12,NMB}.\\
\indent
Another intriguing development, with promising applications to other physical systems,
the anisotropic QRM, interpolating between the QJCM and
the full QRM 
\cite{PhysRevX.4.021046},Tomka},
can be extended to the QRSM 
\cite{QHC,LeiPeter}. Furthermore, asymmetric models have been discussed where the asymmetry term does not break the $\mathbb{Z}_2$ symmetry \cite{Scala}.
Lastly, we mention the polaron picture 
\cite{PhysRevA.92.053823,PhysRevA.95.063803}
which
has been successfully applied to investigate the two-photon 
\cite{PhysRevA.99.013815}
and two-qubit
\cite{SunEtAl}
QRM.
\\
\indent
Currently,
the QRM, and its many variants
are also discussed in connection with the fundamental question of 
quantum entanglement 
\cite{QE,QEP}
which reflects the non-local nature of quantum physics and is, thus, a basic
resource of quantum technology.
Quantum entanglement, on the other hand, is again at the root of such promising
technological developments as quantum computation, 
quantum information 
\cite{Nielsen2010,NakaharaOhmi2008,SuterStolze2004},
which we have already mentioned above, as well as
quantum communication 
\cite{QComm}.\\
\indent
The present study adds to these
developments, especially their theoretical side, 
by studying numerically the phenomenon of 
{\it quantum entanglement resonance} in the asymmetric quantum 
Rabi model (AsymQRM).
We address in particular how the physical quantity of entanglement entropy
can be used to
distinguish level crossings from (narrowly) avoided level crossings. 
This distinction is of central importance in the study of the hidden symmetry of
the asymmetric QRM 
\cite{DB2011,Wakayama2017,SempleKollar2017,Ashhab2020,Mangazeev2021,LiBatchelor2021,Reyes_Bustos_2021,LFTB,PhysRevResearch.3.033057}
where the $\mathbb{Z}_2$ symmetry, and hence level crossings, are restored for
half-integer values of the amplitude of the 
driving term $\epsilon$ in units of the field strength $\omega$.
In order to address this distinction, the numerical accuracy with which spectra 
can be calculated is often insufficient to decide clearly between these two cases,
true level crossings and narrowly avoided level crossings.
The entanglement entropy offers a more sensitive way to distinguish level crossings
from avoided level crossings because it uses not only the eigenvalues but also 
the (low-lying) eigenstates of the Schr\"{o}dinger equation of the AsymQRM.
Studying the von Neumann entanglement entropy 
(for other entanglement entropy notions, see e.g. \cite{BBJS2015})
of the eigenstates of the asymmetric QRM
for different coupling strengths $g$
and driving amplitudes $\epsilon$, we find that the entropy is sensitive to the
spectral structure, exhibiting distinctive resonance valleys when the coupling strength $g$ 
and resonance peaks when the driving amplitude $\epsilon$ is increased.
These resonances occur in both cases at the loci of the avoided level crossings of
the energy spectra.
\\
\indent
We note that the entanglement entropy
has already been used in the study the level crossings in the anisotropic QRM
\cite{PhysRevX.4.021046}, and the
spectral classification of coupling regimes
\cite{Rossatto}, as well as
the QRM in the polaron picture \cite{LiuEtAl}.
Moreover, similar entanglement resonance behaviour has been discussed earlier in
quantum spin chains 
\cite{Karthik}
and periodically driven multi-partite quantum systems
\cite{Sauer}.

%

In Sec \ref{model} of this paper, we give the Hamiltonian and the entanglement characterization 
of the AsymQRM. 
The entanglement resonance with the increase of the photon-atom coupling 
strength is also revealed. 
In Sec. \ref{epsilon}, we study the entanglement resonance with the increase 
of the driving amplitude. 
The entanglement preservation caused by the hidden symmetry is also uncovered. 
Finally, we give a summary in Sec. \ref{conclusion}.
In Appendix \ref{appendix1}, 
we provide some physical intuition
for the entanglement entropy of the simpler asymmetric quantum
Jaynes-Cummings model, where analytical calculations are feasible to a much
greater extent than in the asymmetric quantum Rabi model.

\section{
Entanglement Resonance}\label{model}
The AsymQRM describes the interaction between 
a quantized Bosonic field and a two-level system and is subject to
a static-field driven two-level atom with Hamiltonian
\begin{equation}\label{Hamiltonian}
 \hat{H}=\omega_0\hat{\sigma}_{+}\hat{\sigma}_{-}+\omega \hat{a}^{\dagger}\hat{a}
 +[g (\hat{a}^{\dagger} +\hat{a})+ \epsilon](\hat{\sigma}_{+}+\hat{\sigma}_{-}),
\end{equation}
acting in the tensor Hilbert space ${\mathcal H}_a\otimes{\mathcal H}_f$
where $\hat{a}$ and $\hat{a}^\dagger$ are the Bosonic annihilation and creation
operators acting in the Hilbert space ${\mathcal H}_f$
 with frequency $\omega$, 
$\hat{\sigma}_+=\hat{\sigma}_-^\dag=|e\rangle\langle g|$ are the transition operators 
between the ground state $|g\rangle$ and the excited state $|e\rangle$ 
acting in the two-dimensional Hilbert pace ${\mathcal H}_a$
of the two-level atom with 
frequency $\omega_0$, $g$ is the atom-field coupling strength, and $\epsilon$ is the 
amplitude of static driving.\\
\indent
The atom-field entanglement of any eigenstate $|\Psi\rangle$ of the AsymQRM can be 
quantified by the entanglement entropy which we choose here as
the von Neumann entropy of the 
reduced density matrix for any one of the subsystems \cite{QEP}, 
e.g. the field ($f$) and atomic ($a$) 
subsystems considered here,
\begin{equation}
 S=-\text{Tr}{(\rho_\text{a} \mathrm{log}_2\rho_\text{a})}=-\text{Tr}{(\rho_\text{f} \mathrm{log}_2\rho_\text{f})},\label{SvonNeumannentropy}
\end{equation}
where $\rho_\text{a}=\text{Tr}_\text{f}(|\Psi\rangle\langle\Psi|)$ and 
$\rho_\text{f}=\text{Tr}_\text{a}(|\Psi\rangle\langle\Psi|)$. The entanglement entropy vanishes 
for a separable state and equals one for a maximally entangled state.

\begin{figure}[tbp]
  \includegraphics[width=1\columnwidth]{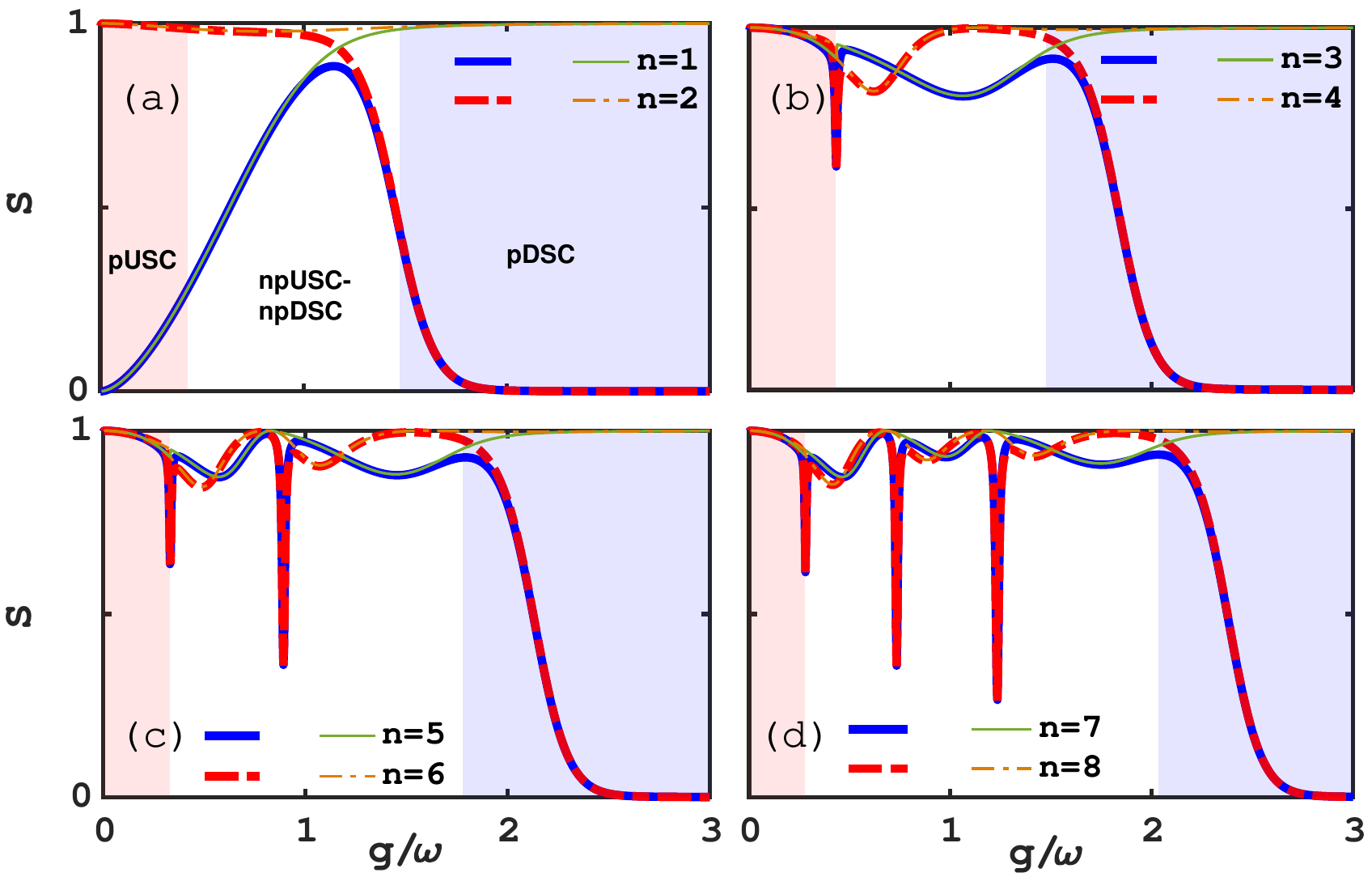}\\
  \caption{Entanglement entropy of different eigenstates $|\Psi_n\rangle$ of the 
  AsymQRM with $\epsilon = 0.01\omega$ denoted by thick lines when $n=1,2$ 
  in (a), $3,4$ in (b), $5,6$ in (c), and $7,8$ in (d) as a function of atom-field 
  coupling strength $g$. 
  The corresponding thin lines denote the results of QRM 
  with $\epsilon=0$. Different colors from left to right denote the pUSC, 
  the npUSC-npDSC, and the pDSC regimes, respectively. 
  We use $\omega_0=\omega$. The truncation number in the numerical calculation is 400.}
  \label{fig1}
\end{figure}

The Hamiltonian \eqref{Hamiltonian} has no obvious symmetry. Therefore the eigenstates $|\Psi_n\rangle$ can only be labeled by the energy
eigenvalues $E_n$ of the Schr{\"{o}}dinger equation
\begin{equation}
H|\Psi_n\rangle=E_n|\Psi_n\rangle.
\end{equation}

We numerically evaluate the entropy expressions 
\eqref{SvonNeumannentropy} by
expanding the Hamiltonian \eqref{Hamiltonian} in the complete basis 
$|m,p\rangle=|m\rangle|p\rangle, m=0,1,2,\ldots, p=\pm$ of the combined 
system of the atom 
and the field 
\begin{equation}
\label{basis}
|\Psi_n\rangle
=\sum_{m,p}c^{(n)}_{m,p}|m,p\rangle
\end{equation}
to obtain the matrix representation 
of the Hamiltonian \eqref{Hamiltonian}
which we use to numerically
obtain eigenvalues and eigenstates in a truncated Hilbert space
at a photon number $n=n_{\mathrm{trunc}}$ such 
that the obtained magnitudes of the eigenenergies converge. \\
\indent
We present in Fig. \ref{fig1} the entanglement entropy of different eigenstates 
$|\Psi_n\rangle$ of the AsymQRM and the QRM 
Hamiltonian, respectively, as a function of the 
atom-field coupling strength $g$. According to the classification rule proposed in Ref. \cite{PhysRevA.96.013849} 
for $\epsilon=0$, the coupling regimes of both, the QRM 
and the AsymQRM, can be divided
into three regions,
the perturbative
ultrastrong coupling (pUSC) regime, 
the perturbative deep strong coupling
(pDSC) regime, and  
the intermediate region, designated as the
nonperturbative ultrastrong-deep 
strong coupling (npUSC-npDSC) regime.  
The regimes pUSC and npUSC-npDSC are separated by the first 
energy-level crossing point and the npUSC-npDSC and pDSC regimes
are separated by the energy-level coalescence point where the adjacent eigenenergies become quasidegenerate. \\
\begin{figure}[tbp]
  \includegraphics[width=1\columnwidth]{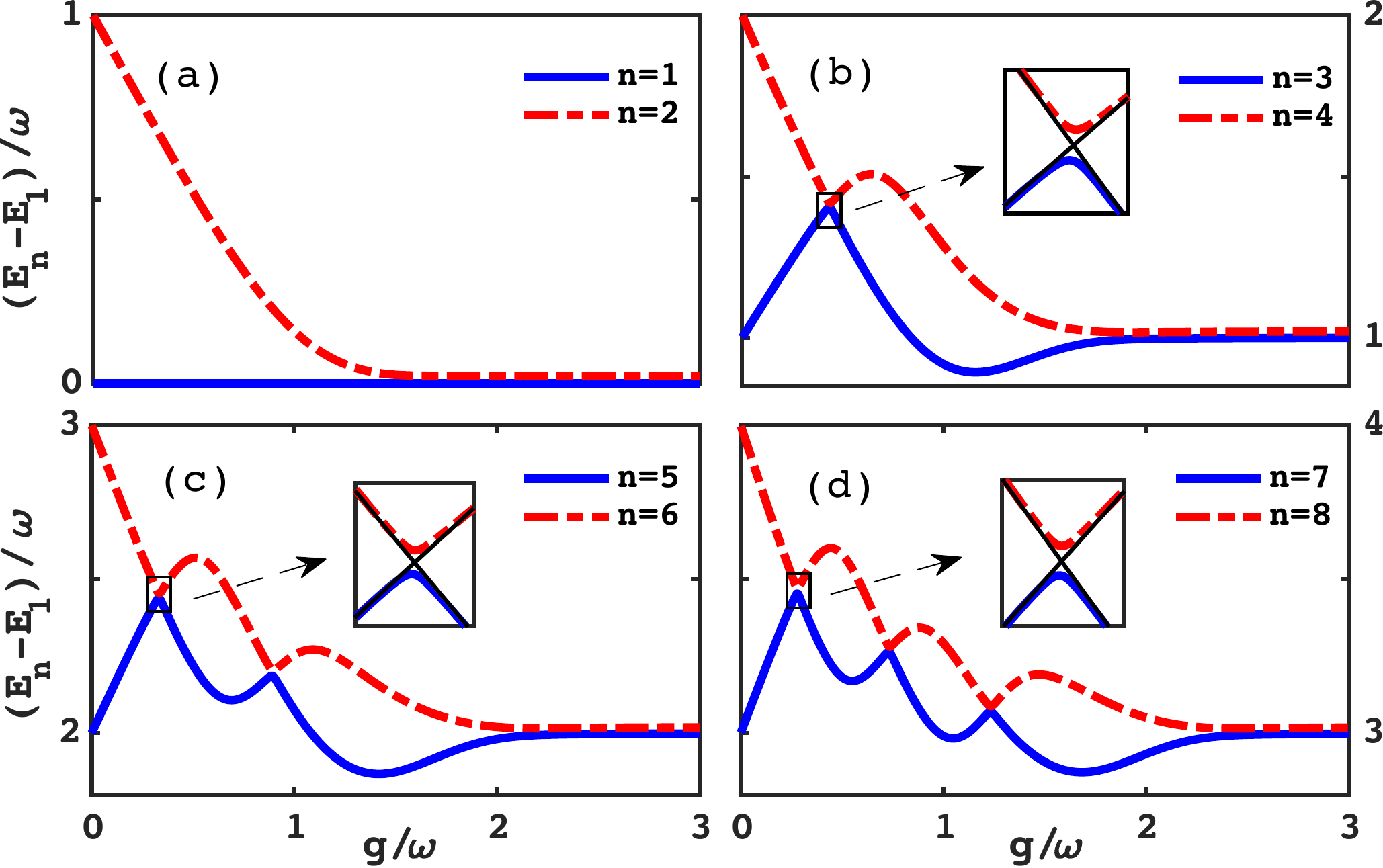}\\
  \caption{Eigenenergies corresponding to the AsymQRM in Fig. 
  \ref{fig1} relative to the ground state energy $E_1$. 
  The gray solid lines in each inset are the eigenenergies of the QRM with $\epsilon=0$. 
  The coupling values of $g$
  of entanglement resonances in the npUSC-npDSC regime in Fig. \ref{fig1} exactly match 
  the ones at which the associated eigenenergies show avoided level crossings.
  The truncation number in the numerical calculation is again 400.}
  \label{avld}
\end{figure}

\indent
This spectral classification is based on the validity of perturbative criteria of the quantum Rabi model, 
which allows the use of exactly solvable effective Hamiltonians.  
When the counter-rotating terms can still be treated perturbatively, 
the system is in the pUSC regime. 
The system is in the pDSC regime, when the interaction term cannot any more be treated as a perturbation but is the main driver of the dynamics.
However, in the  pDSC regime, 
the qubit term $\omega_0$ can be treated perturbatively.
As emphasized in \cite{PhysRevA.96.013849}, this spectral
classification
is a qualitative classification and 
does not imply an abrupt but rather a gradual change in the physical
properties of the QRM and the AsymQRM.\\
\indent
From Fig. \ref{fig1} we infer that, 
in sharp contrast to the case of the
QRM with $\epsilon=0$, the entanglement entropy in the
AsymQRM 
shows a number of resonance valleys in the npUSC-npDSC regime. 
Furthermore the number of the resonance valleys in this weak 
driving case equals to $\lfloor (n-1)/2\rfloor$
(where $\lfloor x\rfloor\equiv\{m\in\mathbb{Z}|m\le x\}$)
matching the number of
energy-level crossing points in the original QRM 
 \cite{DB2011}
 \cite{EJ2017}.
Another remarkable difference of the AsymQRM is that the entanglement entropy in the pDSC regime decays to zero with the increase of the coupling strength $g$, 
while it remains one in the QRM.\\
\indent
The entanglement resonance signifies the efficient coupling of the relevant 
quantum states, which is essentially determined by the energy spectrum of the system. 
To understand the physical reason for the presence of the entanglement resonance 
in the npUSC-npDSC regime, we plot in Fig. \ref{avld} the corresponding 
energy spectrum of the AsymQRM. 
We can see that all the energy-level crossings in the original QRM are opened by 
the static driving in the AsymQRM. 
It is interesting to find that the places of the avoided level crossings in 
Fig. \ref{avld} exactly match the ones of the entanglement
resonances in Fig. \ref{fig1}. 
This can be physically understood as follows. 
The application of the static driving breaks
the $\mathbb{Z}_2$ symmetry
of the original QRM which results in the opening of the energy-level crossings
of the QRM. 
At the points of avoided level crossings, the high mixing of the two associated 
levels with different parities causes an abrupt change to the entanglement of 
the two involved quantum states.

\begin{figure}[tbp]
  \includegraphics[width=1\columnwidth]{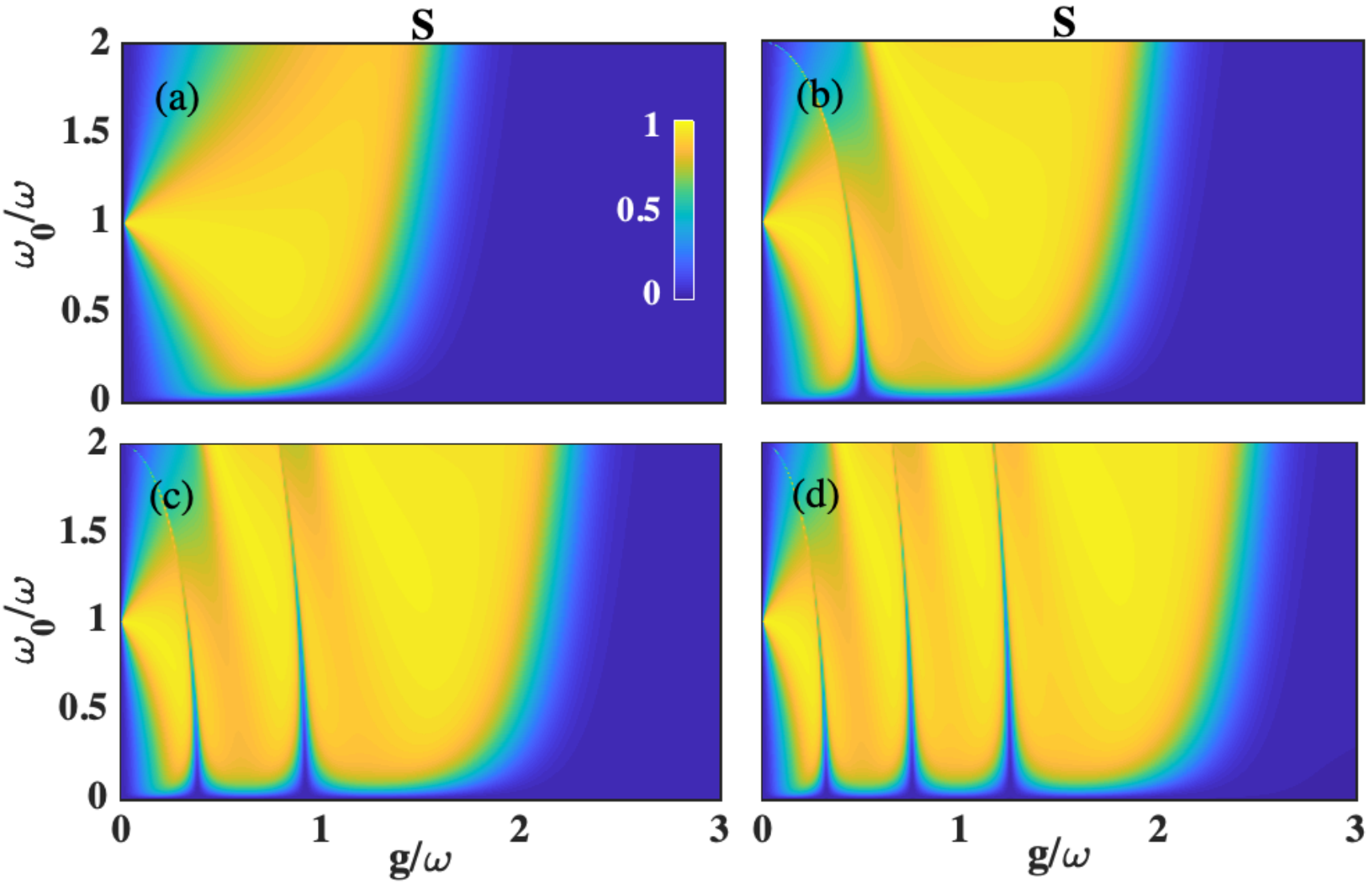}\\
  \caption{Entanglement resonance of the eigenstates $|\Psi_n\rangle$ of the 
  AsymQRM in the parameter plane of $\omega_0$ and $g$ when $n=1$ in (a), 
  $n=3$ in (b), $n=5$ in (c), and $n=7$ in (d). We use $\epsilon=0.1\omega$.}
  \label{entdiag}
\end{figure}

To give a global picture of the entanglement resonance induced by the static driving, 
we plot in Fig. \ref{entdiag} the entanglement entropy in the plane $\omega_0$ versus $g$ for a chosen driving 
amplitude $\epsilon$. 
It can be seen that the entanglement in the weak-coupling limit is almost zero 
except for the resonance case $\omega_0=\omega$. 
Then it changes to one with a tiny increase of $g$. 
With the further increase of $g$ to the npUSC-npDSC regime, 
$\lfloor (n-1)/2\rfloor$ entanglement resonance valleys appear. 
The entanglement in the small-$\omega_0$ limit equals zero. 
It abruptly jumps to one with the increase of $\omega_0$ except for the 
entanglement resonance position. 
Such resonance valleys become sharper and sharper with the increase of $\omega_0$. 
Confirming the entanglement resonance induced by the avoided level crossing, 
the result suggests a useful way to understand the 
features of the energy spectra
of the family of quantum-Rabi models by monitoring 
the entanglement 
between the atom and the quantized field.

\begin{figure}[tbp]
  \includegraphics[width=1\columnwidth]{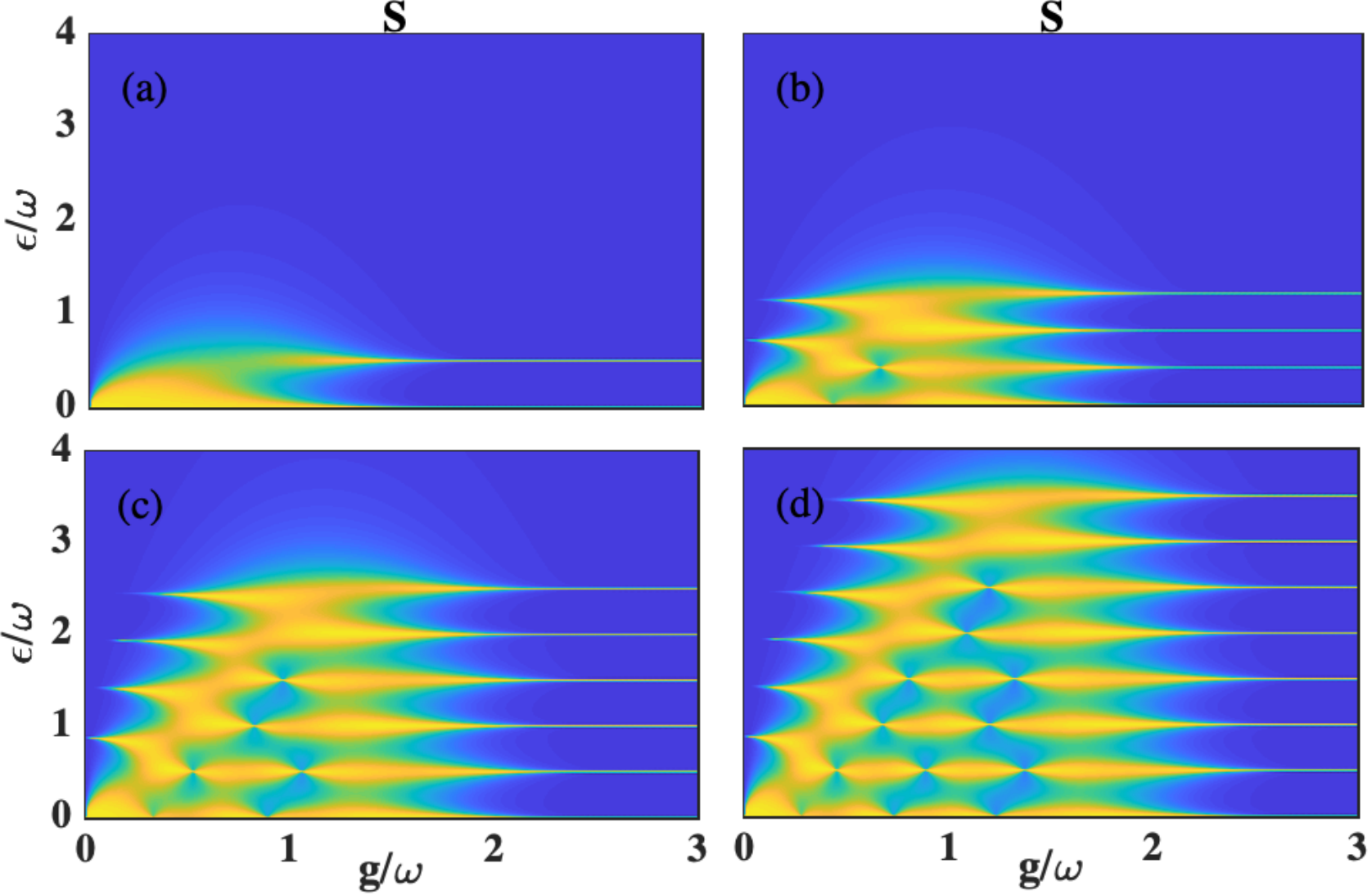}\\
  \caption{Entanglement preservation of the eigenstates $|\Psi_n\rangle$ of the 
  AsymQRM when $\epsilon=m\omega/2$ with $m\in\mathbb{Z}$. 
  Here $n=2$ in (a), $n=4$ in (b), $n=6$ in (c), and $n=8$ in (d). }
  \label{fig4}
\end{figure}

\section{Entanglement preservation in the pDSC regime}\label{epsilon}

We have seen from Fig. \ref{fig1} that the entanglement is not preserved in the 
pDSC regime when a static field is applied. 
Is this valid for arbitrary $\epsilon$? 
To answer this question, we explore the entanglement property of the eigenstates 
$|\Psi_n\rangle$ of the AsymQRM for varying
$\epsilon$.
In Fig. \ref{fig4}, we present the entanglement entropy of $|\Psi_n\rangle$ 
as a function of driving amplitude $\epsilon$ versus
coupling strength $g$ 
for $n=2$, $4$, $6$, and $8$ as examples 
(the result for other, even very large values of n, show the same physics). 
It is revealed that, besides the repeated resonance valleys in the npUSC-npDSC regime,
which have been analyzed in the last section, the entanglement in the pDSC regime 
also shows periodic resonance with increasing $\epsilon$. 
Unlike the resonance valleys with increasing $g$ 
in the npUSC-npDSC regime, 
the entanglement resonance with
increasing $\epsilon$ in the pDSC regime 
shows periodic peaks. 
A maximal entanglement is observed
in the pDSC regime at discrete values 
$\epsilon=m\omega/2$, with $m$ being an integer. 
As $\epsilon$ further increases, the entanglement disappears again.

\begin{figure}[tbp]
  \includegraphics[width=1\columnwidth]{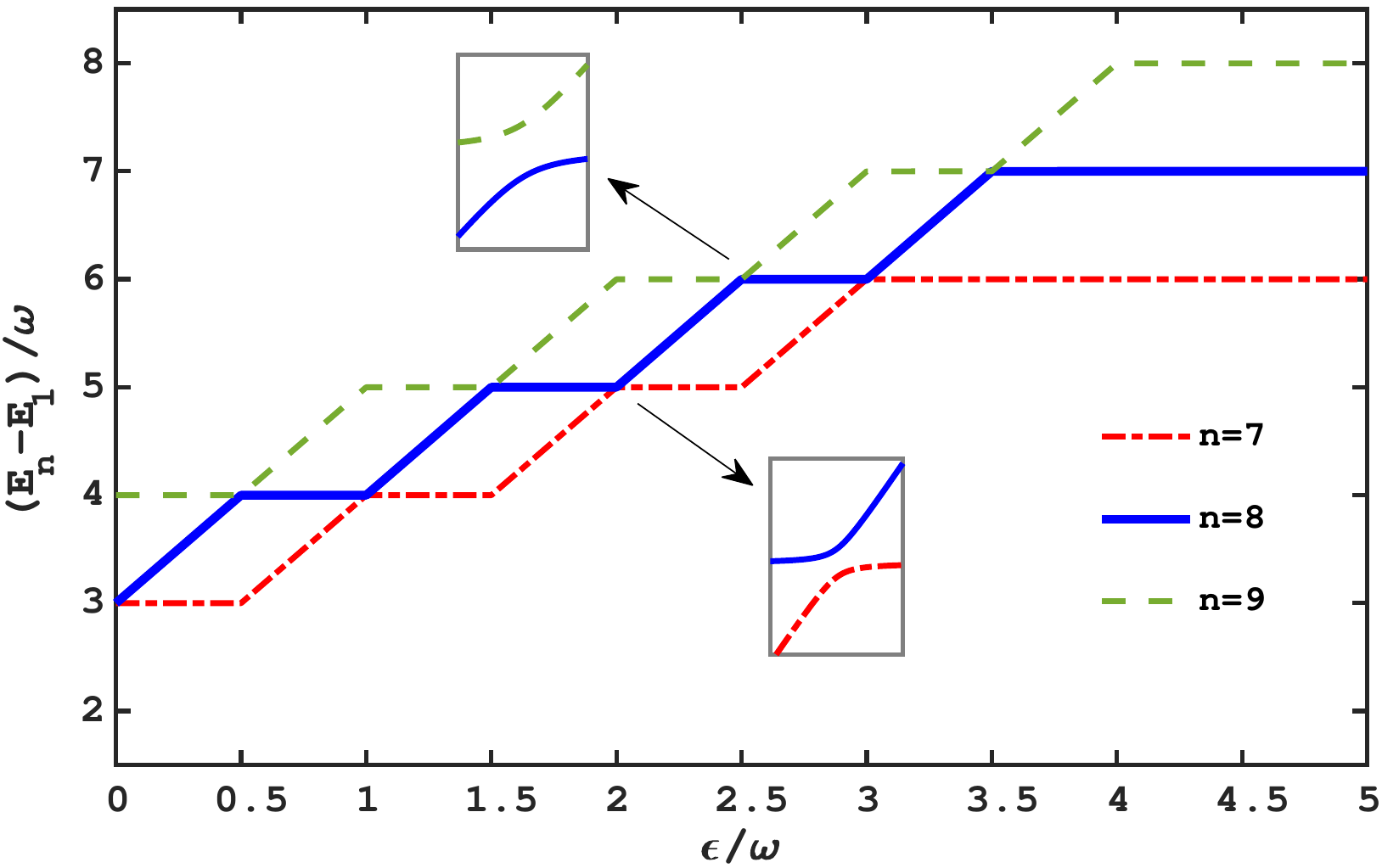}\\
  \caption{Eigenenergies relative to the ground state energy $E_1$ of the AsymQRM as a function of the driving amplitude $\epsilon$ in the pDSC regime when $g=3\omega$. We use $\omega_0=\omega$.  }
  \label{fig5}
\end{figure}

The appearance of the entanglement preservation in the pDSC regime when 
$\epsilon=m\omega/2$ is also caused by the avoided energy-level crossings. 
Taking $n=8$ as an example, we plot in Fig. \ref{fig5} the eighth eigenenergy 
and its nearest-neighbour energies as a function of $\epsilon$ when $g=3\omega$. 
It is interesting to observe that the eighth energy level has eight avoided level 
crossings with its nearest-neighbour levels, which all occur at $\epsilon=m\omega/2$.
These avoided level crossings corresponds exactly with the entanglement preservation in Fig. \ref{fig4}(d). 
The result confirms that the entanglement preservation in the pDSC regime when 
$\epsilon=m\omega/2$ is essentially determined by the avoided level 
crossings.

It was unexpected to observe the restoration, in the pDSC regime of the 
$\mathbb{Z}_2$-symmetry broken AsymQRM, the avoided level crossings, 
also called energy quasi-degeneracies, which are governed by the 
$\mathbb{Z}_2$ symmetry and are present in the QRM. 
Actually, the reappearance of the energy quasi-degeneracies in the AsymQRM 
is caused by a hidden symmetry of the AsymQRM
\cite{Wakayama2017,SempleKollar2017,Ashhab2020,Mangazeev2021,LiBatchelor2021}
occurring 
when $\epsilon=m\omega/2$. Very recently, its symmetry operators for $m=1$ and $m=2$ were rigorously derived
\cite{Mangazeev2021, Reyes_Bustos_2021} and a general scheme to obtain 
the symmetry operators has been proposed \cite{xie2021general}. 
Thus, in addition to the energy-spectrum features which have been investigated 
in previous work, see e.g. \cite{Ashhab2020}, the entanglement 
preservation in the pDSC regime revealed in the present work may
serve as another evidence of the hidden symmetry of the AsymQRM.
\begin{figure}[tbp]
  \includegraphics[width=1\columnwidth]{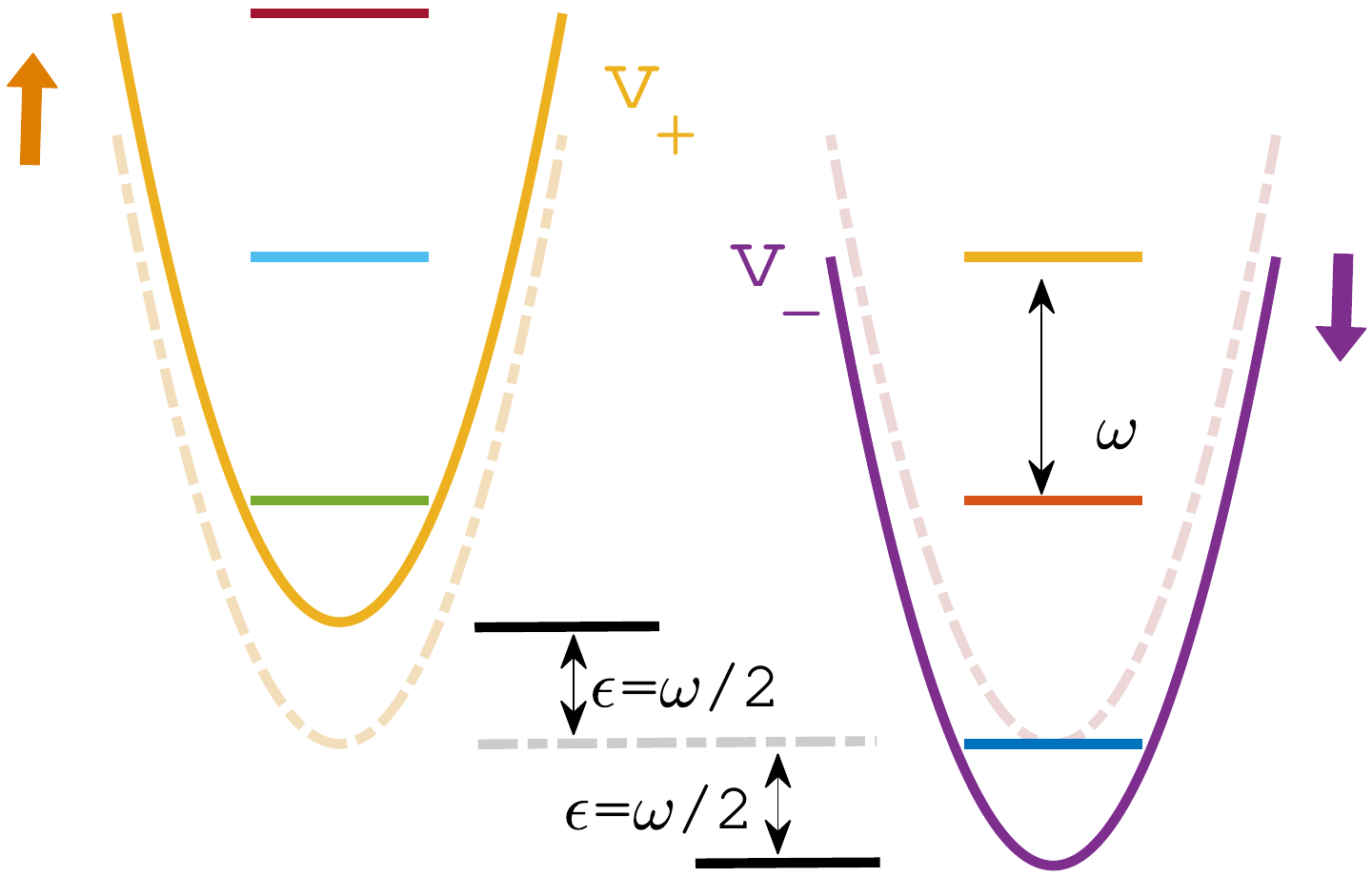}\\
  \caption{Schematic diagram of the two harmonic potentials $V_\pm$ associated to the two $\hat{\sigma}_x$-eigenstates of $\hat{H}_0$. The static driving $\epsilon$ leads to the upward and downward shifts of $V_+$ and $V_-$, respectively. Here $\epsilon=\omega/2$ is shown as an example.}
  \label{fig6}
\end{figure}

Another interesting behavior is that the entanglement preservation occurring at 
$\epsilon=m\omega/2$ only happens for a finite number of integers $m$. 
In order to obtain a physical understanding of this behavior, 
we apply the polaron picture
\cite{PhysRevA.92.053823,PhysRevA.95.063803,PhysRevA.99.013815,SunEtAl}. 
This picture has been applied to a number of variants of the QRM. 
Rotating the Hamiltonian of Eq. (\ref{Hamiltonian}) by the operator 
$e^{(i\pi/4)\sigma_y}$ to
\begin{equation}
\hat{H}= \epsilon\hat{\sigma}_{z}+
\frac{\omega_0}{2}\hat{\sigma}_{x}+
\omega \hat{a}^{\dagger}\hat{a}+g (\hat{a}^{\dagger} +\hat{a})\hat{\sigma}_{z}+
\frac{\omega_0}{2}
\end{equation}
and then
expanding the Hamitonian in the complete basis 
$\sum_{s_x=\pm} |{s_x}\rangle \langle{s_x}|=1$ of $\hat{\sigma}_x\equiv \hat{\sigma}_++
\hat{\sigma}_-$ 
and introducing the unit-mass coordinate 
$\hat{x}=(\hat{a}+\hat{a}^\dag)/\sqrt{2\omega}$ 
and momentum operators
$\hat{p}=i\sqrt{\omega/2}(\hat{a}^\dag-\hat{a})$ 
of the quantized optical field 
\cite{PhysRevA.92.053823,PhysRevA.95.063803}, 
we can rewrite Eq. \eqref{Hamiltonian} as $\hat{H}=\hat{H}_0+\hat{H}_1$ with
\begin{eqnarray}
\hat{H}_0&=&\sum_{s_x=\pm} \hat{h}_{s_x} |{s_x}\rangle \langle{s_x}| +\varepsilon_0,\\
\hat{H}_1&=&\sum_{s_x=\pm} {\omega_0\over2}  |{s_x}\rangle \langle{ \bar{s}_x}| ,\end{eqnarray}
where $\bar{s}_x$ means the flipped spin of $s_x$, 
$\varepsilon_0=(\omega_0-\omega)/2-g^2/\omega$ is a
constant energy, and 
$\hat{h}_{s_x}={\hat{p}^2/2}+\hat{V}_{s_x}$ with
\begin{equation}
\hat{V}_{s_x}={\omega^2\over 2}(\hat{x}+x_{s_x})^2+s_x\epsilon
\end{equation} and $x_{s_x}={\sqrt{2\omega}s_xg/ \omega^2}$. 
Here $\hat{V}_{s_x}$ are harmonic potentials with $s_x=\pm$ labeling the two 
$\hat{\sigma}_x$ eigenstates. In the pDSC regime, taking $\hat{H}_1$ as a perturbation, 
we obtain to leading order the
 eigenenergies of $\hat{H}$ as
\begin{equation}
E^{(0)}_{n,\pm}=n\omega\pm {\epsilon}.
\end{equation}
As illustrated in Fig. \ref{fig6}, one can readily see how the 
driving $\epsilon$ affects the entanglement between the atom and the field. 
The dashed lines represent the case of $\epsilon=0$, where $V_\pm$ are degenerate. 
With increasing
$\epsilon$, $V_+$ shifts upward and $V_-$ shifts downward
with the difference of their valleys being $2\epsilon$. 
The eigenenergies $E^{(0)}_{n,\pm}$ increase or decrease
correspondingly. 
The second energy level {$|2_-\rangle$} (take reference to the notation in \cite{Scala})
in $V_-$ crosses with the second energy level 
$|2_+\rangle$
and the first one$|1_+\rangle$ in $V_+$ 
when $\epsilon=0$ and $\omega/2$, respectively. 
Except for these two values of $\epsilon$, 
$|2_-\rangle$
has no chance to cross with the energy levels in $V_-$ anymore. 
Due to the perturbation of $\hat{H}_1$, such energy-level crossings are reopened, 
which causes a sufficient coupling of $|2_-\rangle$ with $|2_+\rangle$ and $|1_+\rangle$, respectively. 
This generates a large entanglement between the atom and the photon. 
It well explains the result in Fig. \ref{fig4}(a) that a finite 
entanglement for the second 
energy level is preserved in the pDSC regime only when $\epsilon=0$ and $\omega/2$. 
In the same picture, the results in Figs. \ref{fig4}(b), 
\ref{fig4}(c), and \ref{fig4}(d) 
that the entanglement preservation occurs at $\epsilon=m\omega/2$ only 
for $m=0,\cdots, n-1$ can be understood. 
Thus, such a simple picture provides an intuitive explanation 
of the avoided level crossing and entanglement preservation in the pDSC regime.

\section{Conclusions}\label{conclusion}
In summary, we have investigated the entanglement features in the eigenstates 
$|\Psi_n\rangle$ of the coupled system of a quantized optical field with 
a TLS subject to
a statically 
driven two-level atom, i.e., the AsymQRM. 
The entanglement exhibits interesting features depending on the light-matter 
coupling strength and the driving amplitude, which are intrinsically related to the structure of the
energy spectrum of the AsymQRM. We find that the entanglement shows 
$\lfloor (n-1)/2\rfloor$ resonance valleys 
with the change of the light-matter coupling strength in 
the npUSC-npDSC regime. 
Our results indicate that this is caused by the avoided level crossings induced by the 
static field. 
In the stronger pDSC regime, the entanglement exhibits resonance peaks at 
$\epsilon=m\omega/2$, with $m=0,\cdots,n-1$. 
Our analysis demonstrates that such entanglement preservation is induced 
by the avoided level crossings due to the hidden symmetry of the AsymQRM. 
Our result is expected to be helpful to identify the features of the energy spectrum, 
such as the avoided energy level crossings, and to further explore the related 
hidden symmetry properties of more theoretical models for light-matter interaction 
\cite{xie2021double,lu2021hidden,lu2021hidden2}. 
In addition, our method is promising to be applied to study quantum dot(s) with broken inversion symmetry in a cavity
\cite{PhysRevLett.102.023601,PhysRevA.85.053818}.

\section*{Acknowledgements}
We thank Prof. Jun-Hong An for helpful discussions. HPE thanks the School of Physical Science and Technology of 
Lanzhou University for their friendly and supportive hospitality.
This work is supported by the International Postdoctoral Exchange Fellowship 
Program (Grant No. ZD202116), and the National Natural Science 
Foundation of Gansu Province, China (Grant No. 20JR5RA509).

\appendix
\section{Entanglement entropy for the Asymmetric Jaynes-Cummings model}
\label{appendix1}

In order to provide a physical intuition for the 
appearance of the entanglement 
features in the  AsymQRM, we apply a perturbation method to analytically study 
the entanglement entropy in the asymmetric quantum 
Jaynes-Cummings (AsymQJC) model 
which is a model obtained from the
AsymQRM after application of the rotating wave approximation \cite{JC} with the Hamiltonian 

\begin{align}\label{HamiltonianHJC}
\begin{split}
 \hat{H}_\mathrm{AsymQJC}= \frac{\omega_0}{2}\hat{\sigma}_{z}+\omega \hat{a}^{\dagger}\hat{a}+g (\hat{a}^{\dagger}\hat{\sigma}_{-}+\hat{a}\hat{\sigma}_{+})+\epsilon\hat{\sigma}_{x}.
\end{split}
\end{align}

The AsymQJC model shows similar entanglement resonance behavior
as the AsymQRM. For instance, the energy-level crossings in the 
original quantum JC model
are opened by the static driving 
in the AsymQJC  model. 
The quantum JC model is exactly solvable, on
the basis of which, by applying 
the degenerate perturbation method, we obtain the accurate value of the 
entropy for the peak of the entanglement resonance. 

\begin{figure}[tbp]
  \includegraphics[width=1\columnwidth]{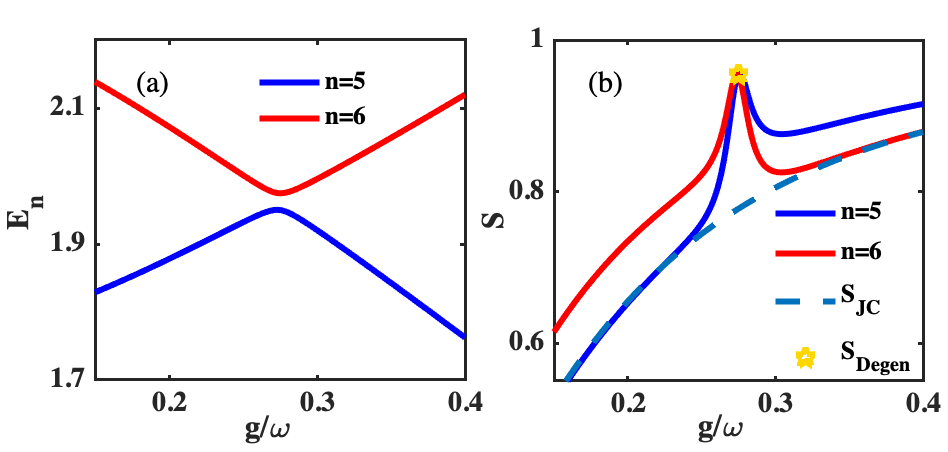}\\
  \caption{(a) Avoided level crossing in the energy spectra of JC model with 
  external perturbation $\epsilon$. 
  (b) Entanglement entropy obtained by numerical (solid lines) and degenerate 
  (yellow star) pertubative method. The entropy of the JC model is also provided 
  as a bench mark. 
  $g/\omega=0.2749$ is the 
  point where the energies of
  the states $|1,+\rangle $ and $|2,-\rangle $ 
  for JC model becomes degenerate.
  $\omega_0=1.5\omega$ and $\epsilon=0.05\omega$.}
  \label{APP1}
\end{figure}

 In Fig. \ref{APP1}, the opened energy gap is presented in
 panel (a) and the numerical 
 and analytical result of the entropy is presented in
 panel (b). 
 In Fig. \ref{APP1} (b), 
 one can see that the analytical solution obtained by the 
 degenerate perturbation 
 method matches the numerical result perfectly. 
 The fact that the entanglement is reduced (or increased) when there is a 
 perturbation that removes a level crossing, can be explained as a consequence 
 of such a simple observation, i.e. if the unperturbed eigenstates are close to 
 being maximally entangled, it is likely that their sum and difference are less entangled. 
 In the same way, if the unperturbed eigenstates are close to being separable, 
 their sum and difference are likely to be highly entangled. 
 The detailed derivation is presented in the
 following section.

\subsection{Analytical solution of the quantum Jaynes-Cummings model}
In this section we summarize the analytical results for the quantum
Jaynes-Cummings model \cite{Book8.10}. The quantum Jaynes-Cummings model with Hamiltonian
\begin{align}
\begin{split}
 \hat{H}_\mathrm{JC}= \frac{\omega_0}{2}\hat{\sigma}_{z}+\omega \hat{a}^{\dagger}\hat{a}
 +g (\hat{a}^{\dagger}\hat{\sigma}_{-}+\hat{a}\hat{\sigma}_{+}).
 \end{split}
\end{align}
is known to be exactly solvable by elementary means.
 It conserves the
total number operator 
$\hat{N}= \hat{a}^{\dagger}\hat{a}+\frac{1}{2}(1+\hat{\sigma}_{z})$.

For $N=0$, the ground state energy is 
$E_{0g}=-\frac{\omega_0}{2}$ and the eigenstate is denoted by
$|g,0\rangle$. 
The excitation energies and the excited states are given by
\begin{equation}\label{JCEnpm0}
E_{n,\pm}=(n+\frac{1}{2})\omega\pm\frac{\Omega_{n,\Delta}}{2},
\quad n=0,1,2,\ldots,
\end{equation}
where $\Omega_{n,\Delta}=\sqrt{\Delta^{2}+4g^2(n+1)}$ and 
$\Delta=\omega_0-\omega$, and
\begin{align}\label{phin}
\begin{split}
 |n,+\rangle&=C_n|n,e\rangle+D_n|n+1,g\rangle,\\
 |n,-\rangle&=D_n|n,e\rangle-C_n|n+1,g\rangle,
\end{split}\end{align}
where $C_n=\cos(\frac{\alpha_n}{2})$ and $D_n=\sin(\frac{\alpha_n}{2})$ with 
$\alpha_n=\tan^{-1}(\frac{2g\sqrt{n+1}}{\Delta})$.





\subsection{Degenerate perturbation for the asymmetric quantum Jaynes-Cummings
model}

The avoided level crossing points in the AsymQJC model correspond
to the doubly 
degenerate, i.e. the crossing, points of the quantum JC model. 
This is the reason we need to apply the 
degenerate perturbation method.

For the Hamiltonian of the quantum JC model at the degenerate point, 
i.e. for a particular value of $g$ to be determined,
the eigenvalue satisfies
\begin{equation}
\mathcal{E}_{n}^{(0)}=E_{n,+}=E_{n+1,-},
\end{equation}
corresponding to two independent and orthogonal 
eigenfunctions $\phi_{n1}=|n,+\rangle$ and $\phi_{n2}=|n+1,-\rangle$,
\begin{equation}\label{HJCwjb}
\hat{H}_\mathrm{JC}\phi_{ni}=\mathcal{E}_{n}^{(0)}\phi_{ni}, \quad i=1,2.
\end{equation}
This eigenvalue, $\mathcal{E}_{n}^{(0)}$, which is exact for the quantum JC model, now
plays the role of the zeroth order approximation eigenvalue for the asymmetric quantum
JC model.

The corresponding zero-order wave function
\begin{equation}\label{HJCwjb}
\varphi_n^{(0)}=\sum_{i=1}^2 c_i^{(0)}\phi_{ni}
\end{equation}
where $c_i^{(0)}$ and the first-order energy eigenvalue $\mathcal{E}_n^{(1)}$ 
can be obtained by solving the eigenvalue equation

\begin{equation}\label{degensilade}
\sum_{j=1}^2(H_{ij}^{'}-\mathcal{E}_n^{(1)}\delta_{ij})c_j^{(0)}=0,
\quad i=1,2
\end{equation}
where $H_{ij}'=\langle\phi_{ni}| \hat{H}' |\phi_{nj} \rangle$ 
represents the coupling between eigenstates $\phi_{ni}$ and $\phi_{nj}$ 
due to the operation of $\hat{H}'=\epsilon\sigma_x$. We {obtain} the first-order modifications of the degenerate eigenenergy
\begin{equation}
\mathcal{E}_{n,\pm}^{(1)}=\pm\epsilon D_nD_{n+1}
\end{equation}
as well as 
$c_1^{(0)}=\pm c_2^{(0)}=\frac{1}{\sqrt{2}}$ for the zeroth-order coefficients
of the eigenstates.

Thus, the total energy at the degenerate point is
\begin{align}\label{degentotalEn}
\begin{split}
E_{n,\pm}=\mathcal{E}_{n}^{(0)}+\mathcal{E}_{n,\pm}^{(1)}.
\end{split}
\end{align}
The modified wave functions $\varphi_{n1}^{(0)}$ and $\varphi_{n2}^{(0)}$ are
\begin{align}\label{degenphi0n12}
\begin{split}
\varphi_{n1}^{(0)}=\frac{1}{\sqrt{2}}(\phi_{n1}+\phi_{n2}),\\
\varphi_{n2}^{(0)}=\frac{1}{\sqrt{2}}(\phi_{n1}-\phi_{n2}).
\end{split}
\end{align}
Note, that the degeneracy of the quantum JC model is now lifted due
to the asymmetry term $\epsilon\sigma_x$, the level crossing at the degenerate point 
of the quantum JC model is now avoided for the asymmetric model.

Finally, we take the wave function $\varphi_{n1}^{(0)}$ as an example to calculate 
its corresponding entanglement entropy. It is known that the total density matrix is 
$\rho_{AB}=|\varphi_{n1}^{(0)}\rangle\langle \varphi_{n1}^{(0)}|$. 
One can get its reduced density matrix $\rho_A\equiv \text{Tr}_{B}(\rho_{AB})$ 
\begin{equation}\label{reduceddensitymatrix}
\rho_A=\left(\begin{array}{ccccc}
 D^2_{n+1}+C^2_n  & D_nD_{n+1} \\
 D_nD_{n+1} & D^2_n+C^{2}_{n+1}  \\
\end{array}\right).
\end{equation}
Now the von Neumann entropy $S_{\rho A}$ of the
zero-level modified wave functions can be obtained
\begin{equation}\label{degenSrho1b}
S_{\rho_A}=-\text{Tr}{[\rho_{A} \mathrm{log}_2(\rho_{A})]}.
\end{equation}

As an example, we calculate the avoided level crossing corresponding to the 
intersection of states $|1,+\rangle$ and $|2,-\rangle$ in the JC model. 
The result is shown in Fig. \ref{APP1}. 
In the case of $\Delta=\omega_0-\omega=0.5\omega$ and $\epsilon=0.05\omega$, 
the crossing point is at $g=0.2749\omega$ 
Then, with the definition of $\alpha_n$, $C_n$ 
and $D_n$ (given below Eq. \ref{phin}), one can calculate the reduced density matrix 
$\rho_A$ and then the entropy $S_{\rho A}$. Note, that the condition for which the 
perturbative method works, $\epsilon \ll \omega,\omega_0,g$, should be satisfied.

\end{document}